\documentclass[%
 reprint,
 amsmath,amssymb,
 aps,
]{revtex4-1}

\usepackage{graphicx}
\usepackage{dcolumn}
\usepackage{bm}

\begin{document}

\preprint{APS/123-QED}

\title{Renormalizability of pure $\mathcal{N}=1$ Super Yang-Mills  in the Wess-Zumino gauge in the presence of the local composite operators $A^{2}$ and $\bar{\lambda}\lambda$\footnote{Error!}}

\author{R.~C.~Terin}
 \email{rodrigoterin3003@gmail.com}
\author{M.~A.~L.~Capri}%
 \email{caprimarcio@gmail.com}
 \author{S.~P.~Sorella}
 \email{silvio.sorella@gmail.com}
 \author{H.~C.~Toledo}
 \email{henriqcouto@gmail.com}
\affiliation{ 
Instituto de F\'isica, Departamento de F\'{\i}sica Te\'{o}rica, 
Universidade do Estado do Rio de Janeiro, Rua S\~{a}o Francisco Xavier, 524, 20550-013, Maracan\~{a}, Rio de Janeiro, RJ, Brasil.
}%


\begin{abstract}
The renormalization of $\mathcal{N}=1$ Super Yang-Mills theory with the presence of the local composite operators $AA$, $A_\mu \gamma_\mu \lambda$ and $\bar{\lambda}\lambda$ is analyzed in the Wess-Zumino gauge, employing the Landau condition. An all-orders proof of the renormalizability of the theory is given by means of the Algebraic Renormalization procedure. Only three renormalization constants are needed, which can be identified with the coupling constant, gauge field, and gluino renormalization. The non-renormalization theorem of the gluon--ghost--anti-ghost vertex in the Landau gauge is shown to remain valid in $\mathcal{N}=1$ Super Yang-Mills with the presence of the local composite operators. Moreover, due to the non-linear realization of the supersymmetry in the Wess-Zumino gauge, the renormalization factor of the gauge field turns out to be different from that of the gluino.
\end{abstract}

\maketitle

First of all, this work is a brief review about what was done in \cite{Capri:2018lru}. If the reader is looking for a more detailed explanations and calculations, we suggest to look the first reference. The supersymmetric gauge theories for $\mathcal{N}=1$ exibits in the perturbative and a non-perturbative regimes some peculiarities, for instance see \cite{Veneziano:1982ah,Amati:1988ft,Gates:1983nr,Shifman:1978bx}.  

In Quantum Field Theory, the expectation values in the vacuum of composite operators play a important role when we want to describe some non-perturbative effects.

In this work, we are going to discuss some renormalization aspects about $\mathcal{N}=1$ Super Yang-Mills, in the Euclidean space-time, with the presence of the local composite operators,$AA$, $A_\mu \gamma_\mu \lambda$ and $\bar{\lambda}\lambda$ in the Wess-Zumino gauge, where the number of field components are minimum. For such proposal, we are going to use the algebraic renormalization procedure \cite{Piguet:1995er}. As well, we are going to show that in the Landau gauge, only the renormalization factors, $(Z_g, Z_A, Z_\lambda)$, identified respectively with the gauge coupling, the gauge field and the gluino field are necessary to renormalizes the theory. Therefore, all the other renormalization factors are written as combinations of $(Z_g, Z_A, Z_\lambda)$. This proposal can be seen as a continuity of other works done in \cite{White:1992ai,Maggiore:1994dw,Maggiore:1994xw,Maggiore:1995gr,
Maggiore:1996gg,Ulker:2001rc,Capri:2014jqa}, which discuss the renormalization of supersymmetric gauge theories through the BRST cohomology. 
 
The proceeding is organized as follows: in Section 1, we perform a brief review on the main aspects of the quantization of the Euclidean super Yang-Mills action in the Wess-Zumino gauge. In Section 2, we discuss the $\mathcal N=1$ Super Yang-Mills quantization in Wess-Zumino gauge with introduction of the $A_{\mu}A_{\mu}$ and $\bar{\lambda}^{\alpha}\lambda_{\alpha}$ operators. In Section 3, we determine the Ward identities and the algebraic characterization of the quantized action. In Section 4, we construct the more general counterterm and we are going to find the renormalization factors for all the fields, external sources and parameters. Finally, we discuss our
conclusions.

\section{ Quantization of $\mathcal N=1$ Euclidean Super Yang-Mills in the Wess-Zumino gauge}
\hspace{0.5 cm} As we have mentioned in the introduction, we are going to use the Wess-Zumino gauge because of the minimum number of components. However, there is a disadvantage: the supersymmetry algebra is realized in a non-linear way. Therefore, the algebra generators of SUSY; $\delta_{\alpha}$, $\alpha=1,2,3,4$, don't close in the translations. Thus,  
\begin{equation} 
\{ \delta_\alpha, \delta_\beta \} = (\gamma_{\mu})_{\alpha\beta} \partial_{\mu} \; + \;\; ({\rm gauge \; transf.}) \; + \;\; ({\rm field \; eqs.}) \;. \label{nlalg}  
\end{equation}
As shown in \cite{White:1992ai,Maggiore:1994dw,Maggiore:1994xw,Maggiore:1995gr,Maggiore:1996gg,Ulker:2001rc}, the most efficient way to handle this kind of algebra, eq.\eqref{nlalg}, is to  construct a generalized BRST operator $Q$ which encodes both susy and gauge transformations, {\it i.e.}
\begin{equation} 
Q=s+\epsilon^{\alpha}\delta_{\alpha} \;, \label{Q}
\end{equation}
where $s$ is the traditional BRST operator for the gauge transformations and $\epsilon^{\alpha}$ is a constant parameter of SUSY, represented by a Majorana spinor, which carries the number of ghost +1. It can also be interpreted as a constant ghost for the SUSY generators. The $Q$ operator has the following property 
\begin{equation} 
Q^{2} = \epsilon^{\alpha}(\gamma_{\mu})_{\alpha\beta}\bar{\epsilon}^{\beta}\partial_{\mu} \;, \label{Q2}
\end{equation}
therefore, we can quantize the theory in order to preserve SUSY invariance through the well known gauge fixing procedure. 

We are going to show how this construction can be applied in a $\mathcal N=1$ Super Yang--Mills theory. The classical action in the Euclidean space reads
\begin{eqnarray}
\label{SYM}
\Sigma_\text{SYM} &=& \int d^{4}x \bigg( \frac{1}{4}F^{a}_{\mu \nu}F^{a}_{\mu\nu} 
+ \frac{1}{2} \bar{\lambda}^{a\alpha} (\gamma_{\mu})_{\alpha\beta} D^{ab}_{\mu}\lambda^{b\beta}
\nonumber \\
&+& \frac{1}{2}\mathfrak{D}^a\mathfrak{D}^a \bigg)\,,  
\end{eqnarray}
an exact $Q$-term.  Adopting the Landau gauge condition, $\partial_\mu A^a_\mu=0$, for the gauge fixing term  we write  
\begin{equation}
\Sigma_\text{gf} = Q\int d^{4}x (\bar{c}^{a}\partial_{\mu}A^{a}_{\mu})\;,  \label{gfx}
\end{equation}
which reads
\begin{equation}
\Sigma_\text{gf} = \int d^{4}x \left[ \bar{c}^{a}\partial_{\mu}D^{ab}_{\mu}c^{b} 
+ b^{a}\partial_{\mu}A^{a}_{\mu} 
- \bar{c}^{a}\bar{\epsilon}^{\alpha}(\gamma_{\mu})_{\alpha\beta}\partial_{\mu}\lambda^{a\beta} \right]\;.  
\label{gfx1}
\end{equation}

\section{ Introducing the local composite operators $A_{\mu}A_{\mu}$ and $\bar{\lambda}^{\alpha}\lambda_{\alpha}$}
\label{sec2}
As written before, we need to describe the set of Ward identities. Then, in order to achieve this aim, we have to add some external sources coupling to non-linear BRST transformations. More precisely, we will couple the some external sources with $QA^a_\mu$, $Q\lambda^{a\beta}$, $Q\mathfrak{D}^a$ and $Qc^a$.  Therefore, we have the following generalized BRST doublets of sources \cite{Piguet:1995er}, namely 
\begin{equation*}
\left\{\begin{matrix}QK^{a}_{\mu}=\Omega^{a}_{\mu}\phantom{\Bigl|}\cr
Q\Omega^{a}_{\mu}=\nabla K^{a}_{\mu}\phantom{\Bigl|}\end{matrix}\right.\,,\qquad
\left\{\begin{matrix}QL^{a}=\Lambda^{a}\phantom{\Bigl|}\cr
Q\Lambda^{a}=\nabla L^{a}\phantom{\Bigl|}\end{matrix}\right.\,,\qquad
\end{equation*}
\begin{equation}
\left\{\begin{matrix}QT^{a}= J^{a}\phantom{\Bigl|}\cr
QJ^{a}=\nabla T^{a}\phantom{\Bigl|}\end{matrix}\right.\,,\qquad
\left\{\begin{matrix}QY^{a\alpha}=X^{a\alpha}\phantom{\Bigl|}\cr
QX^{a\alpha}=\nabla Y^{a\alpha}\phantom{\Bigl|}\end{matrix}\right.\,.
\end{equation}
The corresponding $Q$ invariant external source term is given by 
\begin{equation}
\Sigma_\text{ext} = Q \int d^{4}x \left( -K^{a}_{\mu} A^{a}_{\mu} + L^{a}c^{a} - T^{a} \mathfrak{D}^a  + Y^{a\alpha}\lambda^{a}_{\alpha} \right)\;,  \label{qexact}
\end{equation}
leading to the following $Q$ invariant action $\Sigma_{0}$
\begin{equation}
\Sigma_{0} = \Sigma_{SYM} + \Sigma_\text{gf} + \Sigma_\text{ext}   \;. \label{complact}
\end{equation}
\begin{equation}
Q \Sigma_{0} = 0 \;. 
\end{equation}
Now, we would like to study the model in the presence of the dimension two local gluon and gluino operators $A_{\mu}A_{\mu}$, $A_\mu \gamma_\mu \lambda$ and $\bar{\lambda}^{\alpha}\lambda_{\alpha}$. For this purpose we will employ another set of external sources $(j,\chi,\rho^{\alpha}_{\mu},\tau^{\alpha}_{\mu}, N, R)$, which forms the following doublets of generalized BRST
\begin{equation}
\left\{\begin{matrix}Q \chi = j\phantom{\Bigl|}\cr
Qj = \nabla \chi \phantom{\Bigl|} \end{matrix}\right.\,,\qquad
\left\{\begin{matrix}Q \tau^{\alpha}_{\mu} = \rho^{\alpha}_{\mu} \phantom{\Bigl|}\cr
Q \rho^{\alpha}_{\mu} = \nabla \tau^{\alpha}_{\mu} \phantom{\Bigl|}\end{matrix}\right.\,,\qquad
\left\{\begin{matrix}QR = N \phantom{\Bigl|}\cr
QN = \nabla R \phantom{\Bigl|}\end{matrix}\right.\,,
\end{equation}
and add to eq.\eqref{complact} the following $Q$-invariant term:
\begin{eqnarray}
\Sigma_{AA-\bar{\lambda}\lambda} &=& Q\int d^{4}x\Bigg[\phantom{\Bigl|}\frac{1}{2}\,\chi\, A^{a}_{\mu}A^{a}_{\mu}+\frac{1}{2}\xi\chi j+ \tau^{\alpha}_{\mu}A_{\mu}^{a}\lambda^{a}_{\alpha}\nonumber\\&&+R\bar{\lambda}^{a\alpha}\lambda_{\alpha}^{a} +\frac{\zeta}{4} R N^3 
\Bigg]\,. \label{cop}
\end{eqnarray}
Therefore, for the classical  complete classical action $\Sigma$  including all above-mentioned composite operators, we have
\begin{eqnarray}
\Sigma & = & \int d^{4}x\Bigg[\frac{1}{4}F_{\mu\nu}^{a}F_{\mu\nu}^{a}+\frac{1}{2}\bar{\lambda}^{a\alpha}[\gamma_{\mu}]_{\alpha\beta}D_{\mu}^{ab}\lambda^{b\beta}\nonumber \\
 &  &+\frac{1}{2}\mathfrak{D^{a}}\mathfrak{D^{a}}+\bar{c}^{a}\partial_{\mu}D_{\mu}^{ab}c^{b}+b^{a}\partial_{\mu}A_{\mu}^{a}\nonumber \\
 &  &-\bar{c}^{a}\bar{\epsilon}^{\alpha}[\gamma_{\mu}]_{\alpha\beta}\partial_{\mu}\lambda^{a\beta}+\Lambda^{a}c^{a}
 +L^{a}\big[\frac{1}{2}gf^{abc}c^{b}c^{c} \nonumber \\
 &  &-\bar{\epsilon}^{\alpha}[\gamma_{\mu}]_{\alpha\beta}\epsilon^{\beta}A_{\mu}^{a}\big]-J^{a}\mathfrak{D}^{a} -\Omega_{\mu}^{a}A_{\mu}^{a}+K_{\mu}^{a}\Big[-D_{\mu}^{ab}c^{b}\nonumber \\
 &  &+\bar{\epsilon}^{\alpha}[\gamma_{\mu}]_{\alpha\beta}\lambda^{a\beta}\Big]+T^{a}\big[gf^{abc}c^{b}\mathfrak{D}^{c}\nonumber \\
 &  &-\bar{\epsilon}^{\alpha}[\gamma_{\mu}]_{\alpha\beta}D_{\mu}^{ab}[\gamma_{5}]^{\beta\eta}\lambda_{\eta}^{b}\big] +X^{a\alpha}\lambda_{\alpha}^{a}+gf^{abc}Y^{a\alpha}c^{b}\lambda_{\alpha}^{c}
 \nonumber \\
 &  &+Y^{a\alpha}\big[-\frac{1}{2}(\sigma_{\mu\nu})_{\alpha\beta}\epsilon^{\beta}F_{\mu\nu}^{a}+[\gamma_{5}]_{\alpha\beta}\epsilon^{\beta}\mathfrak{D}^{a}\big]\nonumber \\
 &  &+\rho_{\alpha\mu}A_{\mu}^{a}\lambda^{a\alpha}-\tau_{\alpha\mu}(D_{\mu}^{ab}c^{b})\lambda^{a\alpha}\nonumber \\
 &  &+\tau_{\alpha\mu}\bar{\epsilon}^{\gamma}[\gamma_{\mu}]_{\gamma\beta}\lambda^{a\beta}\lambda^{a\alpha} \nonumber \\
 &  &+gf^{abc}\tau_{\alpha\mu}A_{\mu}^{a}c^{b}\lambda^{c\alpha} -\frac{1}{2}\tau_{\alpha\mu}A_{\mu}^{a}(\sigma_{\rho\nu})^{\alpha\beta}\epsilon_{\beta}F_{\rho\nu}^{a}\nonumber \\
 &  &+\tau_{\alpha\mu}A_{\mu}^{a}[\gamma_{5}]^{\alpha\beta}\epsilon_{\beta}\mathfrak{D}^{a} +\frac{1}{2}jA_{\mu}^{a}A_{\mu}^{a}-\chi A_{\mu}^{a}\partial_{\mu}c^{a}\nonumber \\
 &  &+\chi A_{\mu}^{a}\bar{\epsilon}^{\alpha}[\gamma_{\mu}]_{\alpha\beta}\lambda^{a\beta} +\frac{\xi}{2}j^{2}-\frac{\xi}{2}\chi\bar{\epsilon}^{\alpha}[\gamma_{\mu}]_{\alpha\beta}\epsilon^{\beta}\partial_{\mu}\chi \nonumber \\
 &  &+ N\left(\bar{\lambda}^{a\alpha}\lambda_{\alpha}^{a}\right)+gf^{abc}Rc^{b}\bar{\lambda}^{c\alpha}\lambda_{\alpha}^{a}\nonumber \\
 &  & -\frac{R}{2}(\sigma_{\mu\nu})^{\alpha\gamma}\bar{\epsilon}_{\gamma}F_{\mu\nu}^{a}\lambda_{\alpha}^{a} +R(\gamma_{5})^{\alpha\gamma}\bar{\epsilon}_{\gamma}\mathfrak{D}^{a}\lambda_{\alpha}^{a}\nonumber \\
 &  &-gf^{abc}R\bar{\lambda}^{a\alpha}c^{b}\lambda_{\alpha}^{c} +\frac{R}{2}\bar{\lambda}^{a\alpha}(\sigma_{\mu\nu})_{\alpha\beta}\epsilon^{\beta}F_{\mu\nu}^{a} + \frac{\zeta}{4} N^4\nonumber \\
 &  &-R\bar{\lambda}^{a\alpha}(\gamma_{5})_{\alpha\beta}\epsilon^{\beta}\mathfrak{D}^{a}\Bigg].
\label{SSYM}
\end{eqnarray}
This action will be taken as the starting point  for the algebraic renormalization analysis \cite{Piguet:1995er}.  
\section{Ward identities and algebraic characterization of the invariant counterterm}
The complete action $\Sigma$ obeys a large set of Ward identities, which are describe through in the following: 
\begin{itemize}
{\item The Slavnov-Taylor identity:}
\begin{equation}
\mathcal{S}(\Sigma) = 0 \;, \label{STid}
\end{equation}
where $\mathcal{S}(\Sigma)$ is determined in the reference \cite{Capri:2018lru}.
Let us also mention for later convenience, the so-called linearized Slavnov-Taylor operator $\mathcal{B}_{\Sigma}$ \cite{Piguet:1995er},
which has the following property 
\begin{equation}
\mathcal{B}_{\Sigma}  \mathcal{B}_{\Sigma}  = \nabla \;. \label{bnilp}
\end{equation}
As a consequence, $\mathcal{B}_{\Sigma}$ is nilpotent when acting on space-time integrated local functionals of the fields, sources and their space-time derivatives. For more details see the main reference of this work \cite{Capri:2018lru}.  
\item{The Landau gauge-fixing condition and the  anti-ghost equation \cite{Piguet:1995er}:}
\begin{equation}
\frac{\delta\Sigma}{\delta b^{a}}= i\partial_{\mu}A^{a}_{\mu}\,,\qquad
\frac{\delta\Sigma}{\delta\bar{c}^{a}}+\partial_{\mu}\frac{\delta\Sigma}{\delta K^{a}_{\mu}}=0\,.
\label{GFandAntiGhost}
\end{equation}

\item{The Landau ghost Ward identity  \cite{Piguet:1995er,Blasi:1990xz}:}
\begin{equation}
G^{a}(\Sigma)=\Delta^{a}_{\mathrm{class}}\,,  \label{gW}
\end{equation}
where
\begin{equation}
G^{a}=\int d^{4}x\,\biggl[\frac{\delta}{\delta{c}^{a}} 
+ gf^{abc}\bar{c}^{b}\frac{\delta}{\delta{b}^{c}}\biggl]\,,
\end{equation}
and
\begin{eqnarray}
\Delta^{a}_{\mathrm{class}}&=&\int d^{4}x\,\Bigg[gf^{abc}\bigg(K^{b}_{\mu}A^{c}_{\mu}
-L^{b}c^{c}+T^{b}\mathfrak{D}^{c}
\nonumber\\
&&-Y^{b\alpha}\lambda^{c}_{\alpha}\bigg)-\Lambda^{a}\Bigg]\,. 
 \label{bex}
\end{eqnarray}
It's important to notice that the term $\Delta^{a}_{\mathrm{class}}$, eqs.\eqref{gW},\eqref{bex}, is purely linear in the quantum fields. Other Ward identities are described in \cite{Capri:2018lru}.
\end{itemize}

\section{The algebraic characterization of the invariant counterterm and renormalizability}

Our main purpose is to determine the most general counterterm, for this, we are going to use the well known algebraic renormalization procedure \cite{Piguet:1995er} and we are going to perturb the complete action, $\Sigma$, adding a local integrated polynomial in the fields and in the sources, namely $\Sigma_{count}$, with dimension 4, null number of ghost, and we are going to make that the perturbed action, $(\Sigma + \omega \Sigma_{count})$, obey the same Ward identities of the action $\Sigma$, to the first order in the infinitesimal expansion parameter, $\omega$, in other words,
\begin{equation}
\label{cttnt}
\Sigma_{count}=a_{0}\,S_{\mathrm{SYM}} + \mathcal{B}_{\Sigma} \Delta^{(-1)} \;,
\end{equation}
where $a_0$ is a free coefficient and $\Delta^{(-1)}$ is the most general local integrated polynomial in the fields and external sources with ghost number $-1$ and dimension $3$. 

From the tables displayed in Appendix A of the reference \cite{Capri:2018lru}, the most general expression for $\Delta^{(-1)}$ is written with arbitrary coefficients, namely $a_{i}$, $(i= 1,...,24)$. It's worth to mention that, in according to the tables, we have to choose the dimension $1$ in the ultraviolet for ghost and anti-ghost fields. This choice is a property that simplifies the counterterm analysis, because enable us to determine the dimension ($\frac{1}{2}$), which the supersymmetric parameter has in the ultraviolet.

Applying some symmetries to the counterterm, $\Sigma_{count}$, the following  result is found for $\Delta^{(-1)}$:
\begin{eqnarray}
\Delta^{(-1)} &=& \int d^{4}x \Bigg[a_{1}\Big(\partial_{\mu}\bar{c}^{a} + K^{a}_{\mu}\Big)A^{a}_{\mu}  -\frac{1}{2}a_{1}\Big(\chi A^{a}_{\mu}A^{a}_{\mu}\nonumber \\
&+& 2\tau_{\alpha\mu}A^{a}_{\mu}\lambda^{a\alpha}\Big)+ a_{4}\xi j\chi - \frac{a_{0}}{2}\mathfrak D^{a}T^{a} + a_{11}Y^{a}_{\alpha}\lambda^{a\alpha} \nonumber \\
&+& \Big(\frac{a_{0}}{2} - a_{11}\Big)Y^{a\alpha}[\gamma_{5}]_{\alpha\beta}\epsilon^{\beta}T^{a}+a_{23}R\bar{\lambda}^{a\alpha}\lambda^{a}_{\alpha}
\nonumber\\&+&\frac{a_{24}}{4}RN^{3}\Bigg]\,.\nonumber\\
\label{fct} 
\end{eqnarray}
We observe that $\Sigma_{count}$ contains six arbitrary coefficients $a_{0}$, $a_{1}$, $a_{4}$, $a_{11}$, $a_{23}$  and $a_{24}$, which will give rise to the renormalization factors of all fields, sources and parameters. To complete the algebraic renormalization analysis of the model, we need to demonstrate that the counterterm can be reabsorbed in the starting action through the redefinition of the fields and parameters $\{\varphi \}$, $\varphi = (A,\lambda, b, c, \bar {c}, \mathfrak{D}, \epsilon)$, of the external sources $\{ \Phi \}$, $\Phi= (K,\Omega, \Lambda, T, J, L, Y, X, j,\rho,\tau, N, R)$, of the coupling constant $g$ and of the vacuum parameters $({\xi}, \zeta)$, namely:
\begin{eqnarray}
\label{ration}
\Sigma(\varphi,\Phi,g,\xi) + \omega \Sigma_{count}(\varphi,\Phi,g,\xi)  &=& \Sigma(\varphi_0,\Phi_0,g_0,\xi_{0})\nonumber\\ &+& O(\omega^2) \;, 
\end{eqnarray}
where $(\varphi_0, \Phi_0, g_0,\xi_{0},\zeta_{0})$ stand for  the bare fields, external sources, coupling constant and parameters $(\xi_0,\zeta_0)$. In the present case, we need to take a special care with some mixing of some quantities with the same quantum numbers. In fact, by applying  the linearized Slavnov-Taylor operator to the local integrated polynomial $\Delta^{(-1)}$, we can easily notice that the field $\lambda^{a}$ and the combination $\gamma_{5}\epsilon T^{a}$ have the same dimension and quantum numbers, such as, the $\mathfrak D^{a}$ field and the $Y^{a}\gamma_{5}\epsilon$ combination. In this kind of mixing, we must realize the matrix renormalization, instead, of the multiplicative renormalization. Therefore, the renormalization of $\lambda$ and $\mathfrak{D}$ fields are given by
\begin{equation}
\label{lrenorm}
\lambda^{a\alpha}_{0}=Z^{1/2}_{\lambda}\,\lambda^{a\alpha}+\omega\, z_{1}\,T^{a}(\gamma_{5})^{\alpha\beta}\varepsilon_{\beta}
\end{equation}
and
\begin{equation}
\label{drenorm}
\mathfrak{D}^{a}_{0}=Z^{1/2}_{\mathfrak{D}}\,\mathfrak{D}^{a}+\omega\, z_{2}\,Y^{a\alpha}(\gamma_{5})_{\alpha\beta}\varepsilon^{\beta}\,. 
\end{equation}
Then, we are capable to write each renormalization factor in terms of the coefficients $a_{0}$, $a_{1}$, $a_{4}$, $a_{11}$, $a_{23}$, $a_{24}$, we have 
\begin{eqnarray}
Z_A^{1/2}&=&1+\omega\left(\frac{a_{0}}{2}+a_1\right)\,,\nonumber\\
Z_{g}&=&1-\omega\frac{a_{0}}{2}\,,\nonumber\\
Z_{\lambda}^{1/2}&=&1+\omega\left(\frac{a_{0}}{2}-a_{11}\right)\,,\nonumber\\
Z_{N}&=&(1+\omega a_{23})Z_{\lambda}^{-1}\,, \nonumber \\
Z_{\xi}&=&(1+\omega a_{4})Z_{g}^{-2}Z_{A}\,,\nonumber\\
Z_{\zeta} & = & (1+\omega (a_{24} -a_{23})) Z_{\lambda}^{-4} \;, 
\label{5}
\end{eqnarray}
while, the remaining renormalization factors are combinations of the terms determined in the set of equations described before, for more details see \cite{Capri:2018lru}. This ends the all order algebraic renormalization proof of the supersymmetric Yang-Mills theories  $\mathcal{N}=1$ in presence of the local composite operators $A^2$, $A_\mu \gamma_\mu \lambda$ and $\bar{\lambda}\lambda$. 

Hence, we complete the algebraic renormalization proof to all orders of the supersymmetric Yang-Mills theories  $\mathcal{N}=1$ with the presence of the local composite operators  $A^2$, $A_\mu \gamma_\mu \lambda$ and $\bar{\lambda}\lambda$. Moreover, we have some remarks to do: The six independent parameters $(a_0, a_1, a_4, a_{11}, a_{23}, a_{24})$ are necessary to renormalize the theory. These coefficients corresponds to the
renormalization of coupling constant $g$, to the gauge field $A^a_\mu$, to the parameter $\xi$, to the gluino $\lambda^{a\alpha}$, to the external source $N$ and to the parameter $\zeta$. The other renormalization factors can be written as combinations of $Z_g, Z_A, Z_\lambda$. 

As one can observe form the renormalization factor $Z_N$, which depends from the free parameter $a_{23}$, it does not turn out possible to express the anomalous dimension of the gluino operator $ \bar{\lambda}^{a\alpha}\lambda^{a}_{\alpha}$ in a similar way to that of $A^2$. To some extent, this can be understood as a consequence of the use of the  Wess-Zumino gauge in which supersymmetry and gauge transformations are put together, giving rise to a supersymmetry  algebra which does not close on space-time translations. The same reasoning applies here also to the renormalization factors $Z_A$ and $Z_\lambda$ which turns out in fact to be different from each other, see \cite{Capri:2014jqa} for an explicit higher loop calculations of these  factors. 

\section{Conclusion}

In this work, we realize the renormalization of Super Yang-Mills $\mathcal{N}=1$ model in the Wess-Zumino gauge with the operators $A_{\mu}A_{\mu}$,  $A_\mu \gamma_\mu \lambda$ and $ \bar{\lambda}\lambda$, for this purpose, we use the Landau gauge condition. We followed some authors \cite{White:1992ai,Maggiore:1994dw,Maggiore:1994xw,Maggiore:1995gr,Maggiore:1996gg,Ulker:2001rc,Capri:2014jqa}, to investigate the renormalization of the theory through the BRST symmetry cohomology, thus, we use as a tool the algebraic method \cite{Piguet:1995er}.
Moreover, $Z_{A}$ and $Z_{\lambda}$ are different although both belong to the same multiplet, this is demonstrated in \cite{Capri:2014jqa}. We remember that, the non-renormalization theorem in the Landau gauge of ghost-antighost-gluon vertex \cite{Piguet:1995er,Blasi:1990xz}, in other words, $Z_{c}^{1/2}Z_{\bar{c}}^{1/2}Z_{g}Z_{A}^{1/2}=1$, remains applicable to the supersymmetric version of the theory. Also, we call attention to the renormalization of the source $j$, given by $Z_g^{1/2}Z_{A}^{-1/4}$, which is in complete accordance with the non-supersymmetric case. Such result, being dependent of the renormalization factors $Z_g$ and $Z_{A}$, means that the mass parameter coming from the operator $A^{a}_{\mu}A^{a}_{\mu}$ is not an independent parameter, as expected.
\begin{acknowledgments}
The Conselho Nacional de Desenvolvimento Cient\'{\i}fico e
Tecnol\'{o}gico (CNPq-Brazil), the SR2-UERJ, the
Coordena{\c{c}}{\~{a}}o de Aperfei{\c{c}}oamento de Pessoal de
N{\'{\i}}vel Superior (CAPES)  are gratefully acknowledged.
\end{acknowledgments}


\end{document}